\documentclass[twocolumn,amsmath,prl]{revtex4}
\usepackage{graphicx,bm}
\begin{document}
\title{Oscillations of 2D electron gas photoconductivity
in AC magnetic field}
\author{A. E. Patrakov}
\author{I. I. Lyapilin}\email{Lyapilin@imp.uran.ru}
\affiliation{Institute of Metal
Physics, UD of RAS, Yekaterinburg, Russia}
\begin{abstract}
The response of an electron system to a DC measurement electric field has
been investigated in the case when the system is driven out of the
equilibrium by the magnetic ultra-high frequency field that leads to
combined transitions.
The discussed model includes contributions from Landau quantization and from
microwave irradiation.
Impurity centers are considered as sources of scattering.
It has been shown that the perturbation of the electron system by the
ultra-high frequency magnetic field leads to oscillations of the diagonal
components of the conductivity tensor.
\end{abstract}

\maketitle

There are two main types of spin-orbit coupling in 2D systems based on
semiconductors having the Zincblende structure: Rashba interaction
\cite{rashba}, caused by the structural asymmetry of the quantum well, and
Dresselhaus interaction \cite{dressel}, originating due to the structural
inversion asymmetry of the bulk material.
The spin-orbit interaction (SOI) leads to correlation of the translational and
spin motion of electrons.
AC magnetic field, by acting upon the spin degrees of freedom, indirectly
influences the kinetic degrees of freedom of electrons, thus causing
transitions between energy levels of electrons at combined frequencies.
Thus, such influence would manifest itself also in transport phenomena that
involve only translational degrees of freedom of charge carriers.

We have investigated the response of an electron system to a weak
(``measurement'') DC electric field in the case where the non-equilibrium
state is created by the magnetic ultra-high frequency field that leads to
combined transitions.
The discussed model includes the contributions from Landau quantization and
(in the long-wavelength limit) from the microwave radiation exactly, without
use of the perturbation theory.
We consider impurity centers for the role of scatterers, treating the
scattering process perturbatively.

Since the SOI is in some sense small, then one can perform a
momentum-dependent canonical transformation that decouples kinetic and spin
degrees of freedom.
Naturally, all other terms in the Hamiltonian, describing the interaction
of electrons with the lattice and external fields (if any) also undergo the
transformation.
In this case, the effective interaction of electrons in the system with
external fields appears, which leads to resonant absorption of the field
energy not only at the frequency of the paramagnetic resonance $\omega_s$
or cyclotron resonance $\omega_c$, but also at their linear combinations,
i.e. the combined resonance.

We assume the specific form of the SOI term, namely, Rashba interaction:
\begin{equation}\label{1}
H_{ks}(p) =
\alpha \varepsilon_{zik} \sum_j S^i_j p^k_j =
\frac{i\alpha}{2} \sum_j (S^+_j p^-_j - S^-_j p^+_j),
\end{equation}
$$
S^\pm = S^x \pm i S^y, \quad p^\pm = p^x \pm i p^y.
$$
Here $\alpha$ is the constant characterizing the SOI, $\varepsilon$ is the
fully-antisymmetric Levi---Chivita tensor, $\bm p_j$ and $\bm S_j$ are the
kinetic momentum and the spin of the $j$-th electron.

The effective Hamiltonian obtained after this canonical transformation can
be cast in the form:
\begin{equation}\label{2}
\mathcal{\tilde{H}}(t) = H_0 + H^0_{ef} +
H_{eh}(t) + [T(p), H_{eh}(t) + H^0_{ef} + H_{ev}],
\end{equation}
$$
H_0 = H_k + H_s + H_v + H_{ev}.
$$
Here $H_k$ and $H_s$ are the Hamiltonians representing the kinetic and
Zeeman energy in the magnetic field $\bm H = (0,0,H)$, respectively.
$H^0_{ef}$ is the Hamiltonian of the interaction between the electrons and
the DC electric field $\bm E =(E_x, 0, 0)$.
$H_{eh}(t)$ is the interaction of electrons with the AC magnetic field.
$H_{ev}$, $H_v$ are Hamiltonians of the electron-lattice interaction and of
the lattice itself, respectively.
$T(p)$ is the operator that defines the canonical transformation.
\begin{equation}\label{3}
T(p) = \frac{i \alpha}{2 \hbar(\omega_c-\omega_s)}
\sum_j (S^+_j p^-_j - S^-_j p^+_j).
\end{equation}

The interaction of the spin degrees of freedom of the conductivity electrons
with the AC magnetic field $H_{eh}(t)$ leads to resonant transitions at the
frequency $\omega_s$.
However, as one can see from the expressions above, the effective
interaction $[T(p), H_{eh}(t)]$ leads to combined transitions at frequencies
$\omega_c \pm \omega_s$ and the cyclotron frequency $\omega_c$.
Since, for our further calculations, the response of the non-equilibrium
system to the measurement electric field is interesting, in which the
contribution from the translational degrees of freedom dominates, we will
restrict our consideration to the effective interaction solely.
Besides that, we limit the consideration to the case when the DC and AC
magnetic fields are perpendicular to each other:
$\bm H(t)= (H_x(t),H_y(t),0)$.
In this case, the effective interaction responsible for the combined
transitions has the following form:
\begin{equation}\label{4}
[T(p), H_{eh}(t)]= \frac{i \alpha \omega_{1s}}{2\hbar(\omega_c-\omega_s)}
\sum_j S_j^z ( p_j^+ e^{-i\omega t} - p_j^- e^{i\omega t}),
\end{equation}
$$
\omega_{1s}=\frac{g e H_1}{2 m_0 c},
$$
where $H_1$ is the intensity of the circularly polarized magnetic field,
rotating with the frequency $\omega$.

The dependence of the effective interaction $H_{eh,1}(t)$ upon time causes
certain difficulties while calculating the non-equilibrium response of the
electron system to the measurement electric field.
Thus, it is expedient to carry out one more canonical transformation, that
removes the interaction $H_{eh,1}(t)$ and renormalizes the electron-impurity
interaction Hamiltonian, which acquires the time dependence then.
The renormalized Hamiltonian of the electron-impurity interaction has the
following form:
\begin{multline}\label{5}
\tilde H_{ev}(t)= \sum\limits_{\bm q j, l=-\infty}^{\infty}
V(\bm q) \rho(\bm q) e^{i \bm q \bm r_j} \times{}\\{}\times
\left(\frac{2 S_j^z K_q}{i|K_q|} e^{i\omega t}\right)^l J_l(|K_q|),
\end{multline}
$$
K_q = \frac{\alpha \omega_{1s} ( q_x - i q_y )}
{( \omega_c  - \omega_s ) ( \omega - \omega_c )},
$$
where $V(\bm q)$ is a Fourier component of the potential created by a
single impurity corresponding to the wave vector $\bm q$,
$\rho(\bm q)$ is a Fourier component of the impurity number
density,
and $J_l(x)$ denotes Bessel functions.

In such canonically-transformed system, the impurities act as a coherent
oscillating field that causes resonant transitions.
The physical meaning of the transformation is the change to two
non-uniformly translationally moving reference frames, different for
electrons with opposite spin directions.

The initial non-equilibrium state of the system under consideration is
created by the ultra-high frequency magnetic field and can be described with
the distribution $\bar{\rho}(t)$.
An additional perturbation (e.g., the weak measurement field) leads to the
formation of a new non-equilibrium state.
The task of obtaining the non-equilibrium admittance can be reduced to
finding the transport matrix $T_{BA}(t, \omega)$, which plays in the
non-equilibrium case the same role as in the case of the equilibrium
response.
The real part of the transport matrix determines the relaxation frequency of
the non-equilibrium electrons' momentum.

Assuming the temperatures of the translational and spin subsystems to be
equal (that corresponds to neglecting any heating effects), in the Born
approximation upon the electron-scatterer interaction we obtain for the
relaxation frequency:
\begin{multline}\label{6}
\frac{1}{\tau} = \frac{1}{2 m n T}
\operatorname{Re} \frac{1}{i\hbar}
\int\limits_{-\infty}^0 dt_1 e^{(\varepsilon - \omega_1)t_1}
\int\limits_{-\infty}^0 dt_2 e^{\varepsilon t_2}
\int\limits_0^1 d\lambda
\times{} \\ {}\times
\operatorname{Sp}\{
\dot P^+_{(\tilde v)}(t) e^{i L_0 (t_1 + t_2)}
\rho_q^\lambda \times{}\\{}\times
[\dot P^-_{(\tilde v)}(t+t_1+t_2), H_k + H_s] \rho_q^{1-\lambda}
\},
\end{multline}
where
$$
e^{i L_0 t} A=e^{-\frac{i H_0 t}{\hbar}} A e^{\frac{i H_0 t}{\hbar}},\quad
\dot{P}^\pm_{(\tilde v)}=\frac{1}{i\hbar}[P^\pm, \tilde H_{ev}],
$$
$m$, $n$, and $T$ are the effective mass, the electron density and the
temperature expressed in the units of energy, $\rho_q = \exp(-S) $ is the
quasiequilibrium statistical operator, and $S$ is the entropy operator.

Now we perform the integration upon $\lambda$, $t_1$ in (\ref{6})
and carry out the necessary averaging. For the zero-frequency
response ($\omega_1 \to 0$), the radiation-induced contribution to
the inverse relaxation time can be written as:
\begin{multline}\label{7}
\Delta(\frac{1}{\tau}) = -\frac{\pi\hbar}{2 m n}
\sum\limits_{\bm q \mu \nu l} \int d\mathcal E
|V(q)|^2 N_i J_l^2(|K_q|) q^2
\times{} \\ {} \times
|(2 S^z e^{i \bm q \bm r})_{\nu\mu}|^2
(f(\mathcal E + l \hbar\omega) - f(\mathcal E))
\times{} \\ {} \times
\delta(\mathcal E - \varepsilon_\mu)
\frac{\partial}{\partial\mathcal E}
\delta(l\hbar\omega + \mathcal E - \varepsilon_\nu),
\end{multline}
where $\mu$, $\nu$ denote electron states in magnetic field,
characterized by the Landau level number, the wave vector in the
$x$ direction, and the spin direction: $|\mu\rangle = |n_\mu,
k^x_\mu, S^z_\mu\rangle$. $f(\varepsilon)$ is the Fermi---Dirac
distribution function. $N_i$ is the impurity concentration.

The singularity in the right hand side of (\ref{7}) is removed, as usual,
due to broadening of the Landau levels by scattering electrons on
impurities:
\begin{equation}\label{8}
\delta(\mathcal E - \varepsilon_\mu) \to D_\mu(\mathcal E) =
\frac{\sqrt{\pi/2}}{\Gamma}
\exp\left(-\frac{(\mathcal E - \varepsilon_\mu)}{2 \Gamma^2}\right).
\end{equation}
The Landau level width $\Gamma$ can be expressed via the electron mobility $\mu$
in zero magnetic field.
For the case of point scatterers, when $V(q)$ doesn't depend on $q$, the
correction to the inverse relaxation time can be expressed as:
\begin{multline}\label{9}
\Delta(\frac{1}{\tau}) = \frac{\hbar}{4 m n \ell^8}
\sum\limits_{ n_\nu n_\mu l=\pm 1 }
\frac{|V(q)|^2 N_i\alpha^2 \omega_{1s}^2}
{(\omega_c - \omega_s)^2(\omega - \omega_c)^2}
\times{} \\ {}
\times(n_\nu^2 + n_\mu^2 + 3 (n_\nu + n_\mu) + 4 n_\nu n_\mu + 2)
\times{}\\{}\times
(f(\varepsilon_\nu)-f(\varepsilon_\mu))
\frac{\pi^{1/2}(\varepsilon_\mu - \varepsilon_\nu + l
\hbar\omega)}{\Gamma^3}
\times{}\\{}\times
\exp\left(
-\frac{(\varepsilon_\mu - \varepsilon_\nu + l \hbar\omega)^2}
{4\Gamma^2}
\right),
\end{multline}
where $\ell = \sqrt{c \hbar / (|e| H)}$ is the magnetic length.

\begin{figure}
\center\includegraphics[width=8cm]{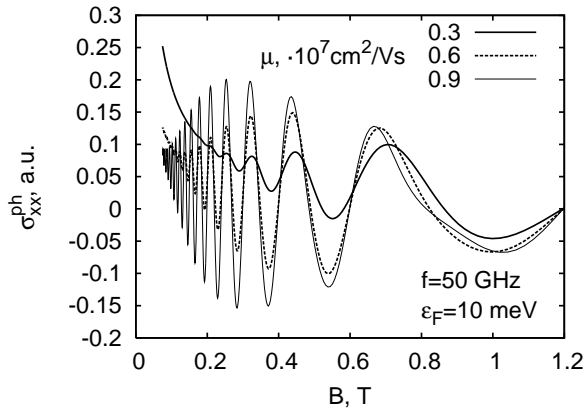}
\caption{Photoconductivity of the 2D electron gas as a function of the
magnetic field induction for different values of electron mobility. The
radiation frequency is 50~GHz and $\gamma=2$.}\label{fig:Lmuhmu}
\end{figure}

Using the expression for the momentum relaxation rate, one can also
calculate the diagonal components of the conductivity tensor $\sigma_{x x}$.
Numerical calculations of these components have been carried out with the
following parameters:
$m = 0.067 m_0$ ($m_0$ is the free electron mass),
the Fermi energy is $\mathcal{E}_F = 10$~meV,
the mobility of the 2D electrons varies as
$\mu\approx 0.1 - 1.0\cdot 10^7$~cm$^2$/Vs,
the electron density $n=3\cdot 10^{11}$~cm$^{-2}$.
The microwave radiation frequency is $f= 50$~GHz,
the temperature is $T\approx 2.4$~K.
The magnetic field varied as 0.02~--~0.3~T.
The results of the numerical calculation are presented in
Fig.~\ref{fig:Lmuhmu}.

It follows from the analysis that the dependence of the electron mobility
upon the magnetic field is oscillatory.
In the region of weak magnetic fields, the amplitude of the oscillations is
very sensitive to the width of Landau levels and decreases noticeably with
the decrease of the zero-magnetic-field electron mobility.

{\em In conclusion}, the response of a non-equilibrium electron system to the DC
electric measurement field has been studied for the case when the initial
non-equilibrium state of the system is created by an ultra-high frequency
magnetic field.
It has been shown that such perturbation of the electron system essentially
influences the transport coefficients and leads to the oscillations of the
diagonal components of the conductivity tensor.
The discussed effect is analogous to the phenomenon observed in
GaAs/AlGaAs heterostructures with ultra-high electron mobility \cite{Mani}.
However, unlike that phenomenon, the manifestation of the oscillatory
pattern is dictated by the spin-orbit interaction existing in the crystals
under consideration.

\end{document}